\documentclass[
	aps,prb,
	superscriptaddress,
	eqsecnum,
	twocolumn,
	amsmath,amssymbol,
]{revtex4-1}

\usepackage{graphicx}

\begin{document}


	\title{Successive magnetic transitions relating to itinerant spins and localized Cu spins in La$_{2-x}$Sr$_x$Cu$_{1-y}$Fe$_y$O$_4$: Possible existence of stripe correlations in the overdoped regime}

	\author{Kensuke M. Suzuki}
		\altaffiliation[Present address: ]{Institute for Materials Research, Tohoku University, 2-1-1 Katahira, Aoba-ku, Sendai 980-8577, Japan.}
		\affiliation{Department of Applied Physics, Graduate School of Engineering, Tohoku University, 6-6-05 Aoba, Aramaki, Aoba-ku, Sendai 980-8579, Japan}
	\author{Tadashi Adachi}
		\email[Corresponding author: ]{t-adachi@sophia.ac.jp}
		\affiliation{Department of Engineering and Applied Sciences, Faculty of Science and Technology, Sophia University, 7-1 Kioi-cho, Chiyoda-ku, Tokyo 102-8554, Japan.}
	\author{Hidetaka Sato}
		\affiliation{Department of Applied Physics, Graduate School of Engineering, Tohoku University, 6-6-05 Aoba, Aramaki, Aoba-ku, Sendai 980-8579, Japan}
	\author{Isao Watanabe}
		\affiliation{Advanced Meson Science Laboratory, Nishina Center for Accelerator-Based Science, The Institute of Physical and Chemical Research (RIKEN), 2-1 Hirosawa, Wako 351-0198, Japan}
	\author{Yoji Koike}
		\affiliation{Department of Applied Physics, Graduate School of Engineering, Tohoku University, 6-6-05 Aoba, Aramaki, Aoba-ku, Sendai 980-8579, Japan}

	\date{\today}

	\begin{abstract}
			In order to investigate the relationship between the so-called stripe correlations and the high-$T_\mathrm{c}$ superconductivity,
			we have carried out magnetic susceptibility and zero-field muon-spin-relaxation measurements in partially Fe-substituted La$_{2-x}$Sr$_x$Cu$_{1-y}$Fe$_y$O$_4$ (LSCFO).
			It has been found that the Fe substitution induces successive magnetic transitions in the overdoped regime, namely, a spin-glass transition of Fe$^{3+}$ spins due to the RKKY interaction based on itinerant spins at higher temperatures and a stripe-order transition of localized Cu$^{2+}$ spins at lower temperatures.
			The stripe-order transition disappears at the hole concentration of $\sim 0.28$ per Cu which coincides with the endpoint of the superconductivity in pristine La$_{2-x}$Sr$_x$CuO$_4$, suggesting that the stripe correlations are closely related to the high-$T_\mathrm{c}$ superconductivity in the overdoped regime as well as in the underdoped regime.
			As for the spin-glass state, it has been found that the contribution of polarized itinerant spins to the muon-spin depolarization is not negligible. Taking into account neutron-scattering results by R.-H. He \textit{et al.} [Phys. Rev. Lett. {\textbf 107}, 127002 (2011)] that a spin-density-wave (SDW) order has been observed in overdoped LSCFO, the present results demonstrate the presence of a novel coexistent state of a SDW order and a spin-glass state.
	\end{abstract}

\maketitle

\section{Introduction}\label{sec:introduction}
			The so-called stripe correlations of holes and spins\cite{Tranquada1995} have been drawing intense interest in the research field of the high-$T_\mathrm{c}$ cuprate superconductivity.
			According to the theory by S. A. Kivelson \textit{et al.},\cite{Kivelson2003} the stripe correlations can be a glue to form electron pairs in the high-$T_\mathrm{c}$ cuprates.
			Comprehensive experiments of neutron scattering,\cite{Yamada1998,Fujita2002,Fujita2004} muon spin relaxation ($\mu$SR),\cite{Niedermayer1998,Watanabe2002,Adachi2004Zn,Adachi2008} nuclear magnetic resonance\cite{Watanabe1990,Mahajan1994,Julien1999a} and so on have uncovered possible existence of the stripe correlations in the underdoped and optimally doped cuprates.
			The intimate relation between the so-called pseudogap and the stripe correlations has been reported from the scanning tunneling microscope study.\cite{Parker2010a}
			Moreover, it has been suggested from transport measurements that the quantum critical point of the stripe correlations is located at the hole concentration per Cu, $p$, of $\sim 0.24$ in La$_{1.6-x}$Nd$_{0.4}$Sr$_x$CuO$_4$.\cite{Daou2008}
			As for the overdoped regime where the stripe correlations become weak due to the overdoping of holes, the inelastic neutron-scattering experiment has revealed that the incommensurate spin-correlation characteristic of the stripe correlations persists up to $p \sim 0.30$ in La$_{2-x}$Sr$_x$CuO$_4$ (LSCO) at which the superconductivity disappears.\cite{Wakimoto2007}
			Since the impurity substitution often operates to slow down various kinds of fluctuations and investigate the relatively-low-energy excitations, partially Zn-substituted La$_{2-x}$Sr$_x$Cu$_{1-y}$Zn$_y$O$_4$ (LSCZO) has been prepared and the $\mu$SR measurements have been carried out by Risdiana \textit{et al.}\cite{Risdiana2008}
			As a result, it has also been suggested that the stripe correlations are developed up to $p \sim 0.30$,\cite{Risdiana2008} though it has been rather hard to draw a final conclusion on account of the weakness of the spin correlation developed by the Zn substitution.

			Recently, the elastic neutron-scattering experiment by M. Fujita \textit{et al.}\cite{Fujita2008} has revealed that the stripe correlations are much enhanced by the substitution of a small amount of Fe$^{3+}$ for Cu$^{2+}$ in La$_{2-x}$Sr$_x$Cu$_{1-y}$Fe$_y$O$_4$ (LSCFO) with $p\ (\equiv x - y) = 0.12$.
			Even in the overdoped regime, moreover, the Fe substitution has been found to induce clear incommensurate magnetic peaks in the elastic neutron-scattering experiment, so that the magnetic order has been suggested to be a spin-density-wave (SDW) order driven by the Fermi-surface nesting.\cite{He2011PRL}
			From our former $\mu$SR measurements in 1\% Fe-substituted LSCFO, it has been found that a muon-spin precession due to the formation of a long-range stripe order is clearly observed in the underdoped regime, while a fast muon-spin-depolarization without any muon-spin precession probably due to the formation of a spin-glass state of Fe$^{3+}$ spins mediated by the Ruderman-Kittel-Kasuya-Yosida (RKKY) interaction is observed in the overdoped regime.\cite{KMS2012proc,KMS2012}
			In the overdoped regime of partially Fe-substituted Bi$_{1.75}$Pb$_{0.35}$Sr$_{1.90}$Cu$_{0.91}$Fe$_{0.09}$O$_{6+\delta}$ also, the elastic neutron-scattering experiment has revealed that incommensurate magnetic peaks are induced by the Fe substitution.\cite{Hiraka2010,Wakimoto2010}
			The relationship between the magnetism in the overdoped regime induced by the Fe substitution and the stripe correlations attracts great interest. Therefore, in order to investigate the relationship, we have carried out magnetic susceptibility and zero-field (ZF) $\mu$SR measurements in LSCFO changing $p$ and the amount of substituted Fe widely.
			
\section{Experimental}
			Polycrystalline samples of LSCFO with $x = 0.10$ -- $0.30$ and $y = 0.01$ -- $0.10$ were prepared by the ordinary solid-state reaction method.\cite{Adachi2004Zn}
			All the samples were checked by the powder x-ray diffraction to be of the single phase.
			Electrical resistivity measurements have revealed the good quality of the samples.
			Magnetic susceptibility measurements were performed at temperatures down to 2 K, using a superconducting (SC) quantum interference device magnetometer (Quantum Design, MPMS-XL7).
			ZF-$\mu$SR measurements were performed using the GPS spectrometer and the dc positive surface muon beam at the Paul Scherrer Institute in Switzerland.
			The asymmetry parameter $A(t)$ at a time $t$ was given by
			$A(t) = \{ F(t) - \alpha B(t)\}/\{ F(t) + \alpha B(t)\}$,
			where $F(t)$ and $B(t)$ were total muon events of the forward and backward counters, which were aligned in the beam line, respectively.
			The $\alpha$ is the calibration factor reflecting  the relative counting efficiencies between the forward and backward counters.
			The $\mu$SR time spectrum, namely, the time evolution of $A(t)$ was measured at low temperatures down to 1.6 K.
			The data were analyzed using the WiMDA program.\cite{Pratt2000}

\section{Results and Discussion}
	\subsection{Fe-concentration dependence of the magnetic susceptibility and $\mu$SR in overdoped La$_{2-x}$Sr$_x$Cu$_{1-y}$Fe$_y$O$_4$ with $p = 0.20$}\label{sec:RAD1}

			\begin{figure}[t]
				\includegraphics[width=0.85\linewidth]{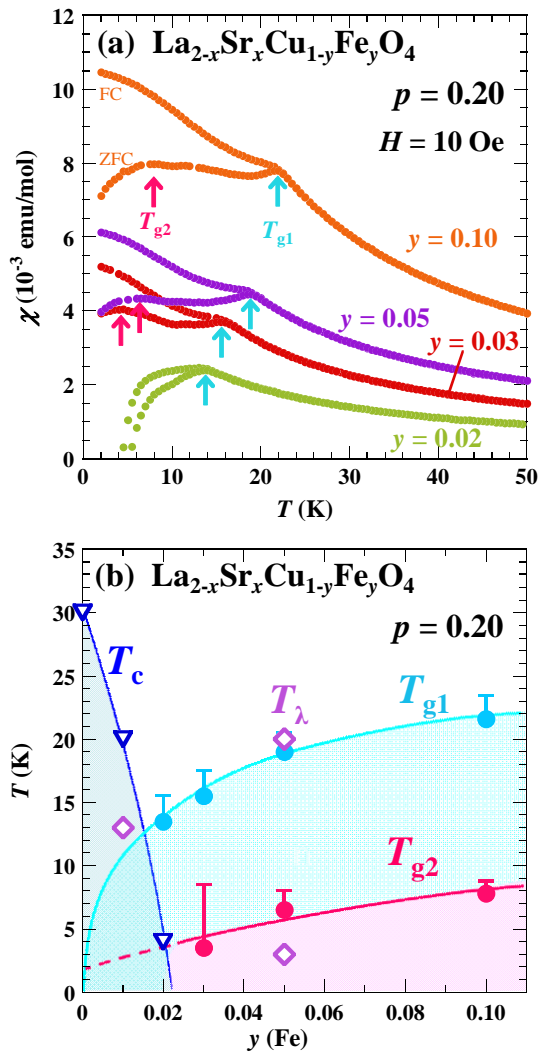}
				\caption{\label{chiT0.20}
					(color online)
					(a) Temperature dependence of the field-cooled (FC) and zero-field-cooled (ZFC) magnetic susceptibility in a magnetic field of 10 Oe for La$_{2-x}$Sr$_x$Cu$_{1-y}$Fe$_y$O$_4$ (LSCFO) with the hole-concentration per Cu, $p\ (\equiv x - y), = 0.20$ and $y = 0.02$ -- $0.10$ .
					The arrows represent $T_\mathrm{g1}$ or $T_\mathrm{g2}$ defined as the temperature at which the ZFC data show a local maximum.
					(b) Fe-concentration dependences of $T_\mathrm{g1}$, $T_\mathrm{g2}$, $T_{\lambda}$ and the superconducting transition temperature, $T_\mathrm{c}$, for LSCFO with $p = 0.20$.\cite{KMS2012,Koike1992}
					Error bars represent the inflection temperature below which the ZFC magnetic susceptibility begins to be suppressed downward.
					The $T_{\lambda}$ is defined as the temperature at which the depolarization rate of the first component in Eq. (\ref{eq1}), $\lambda _0$, of $\mu$SR shows a local maximum.
					Solid and dashed lines are guides to the eye.
				}
			\end{figure}
		
			Figure \ref{chiT0.20}(a) shows the temperature dependence of the field-cooled (FC) and zero-field-cooled (ZFC) magnetic susceptibility in a magnetic field of 10 Oe in overdoped LSCFO with $p = 0.20$ and $y = 0.02$ -- $0.10$.
			The superconductivity is completely suppressed by the Fe substitution except for $y = 0.02$ where the onset of the SC transition is observed at $\sim 7$ K.
			A Curie-Weiss-like behavior is observed at high temperatures for all the samples, indicating a paramagnetic state of Fe$^{3+}$ spins.
			With decreasing temperature, it is found that all the data deviate downward from the Curie-Weiss-like behavior and show a hysteresis between the FC and ZFC data accompanied by a local maximum in the ZFC data.
			These behaviors indicate the formation of a spin-glass state.
			Surprising is that another local maximum is observed at a lower temperature in the ZFC data of $y = 0.03$ -- $0.10$, suggesting the appearance of another magnetic order.
		
			
			The higher (lower) temperature at which the magnetic susceptibility shows a local maximum is defined as $T_\mathrm{g1}$ ($T_\mathrm{g2}$).
			Fe-concentration dependences of $T_\mathrm{g1}$, $T_\mathrm{g2}$ and the SC transition temperature, $T_\mathrm{c}$, are shown in Fig. \ref{chiT0.20}(b).\cite{KMS2012,Koike1992}
			It is found that both $T_\mathrm{g1}$ and $T_\mathrm{g2}$ increase progressively with increasing $y$, suggesting that the two magnetic phases are developed by the Fe substitution.
			Considering that LSCFO with $x = 0.21$ and $y = 0.01$ ($p = 0.20$) undergoes a spin-glass transition due to the RKKY interaction at 13 K as reported in our previous paper\cite{KMS2012} (shown by $\diamond$ in Fig. \ref{chiT0.20}(b)) and that the extrapolation of $T_\mathrm{g1}$ into the region below $y = 0.02$ passes close to the magnetic transition temperature of 13 K at $y = 0.01$, the spin-glass state between $T_\mathrm{g1}$ and $T_\mathrm{g2}$ is of Fe$^{3+}$ spins due to the RKKY interaction.
			The extrapolation of $T_\mathrm{g1}$ also suggests the coexistence of the superconductivity with the spin-glass state at low temperatures below $T_\mathrm{g1}$ for $y < 0.02$.
			As for the magnetic order developed at low temperatures below $T_\mathrm{g2}$, it also appears to persist in the SC phase even at $y = 0$, considering the onset temperature at which the magnetic susceptibility begins to be suppressed downward as shown by the upper limit of the error bar in Fig. \ref{chiT0.20}(b).
			However, the magnetic transition at $T_\mathrm{g2}$ may be suppressed by the superconductivity for $y \lesssim 0.02$.

		
		\begin{figure*}[t]
			\includegraphics[width=0.8\linewidth]{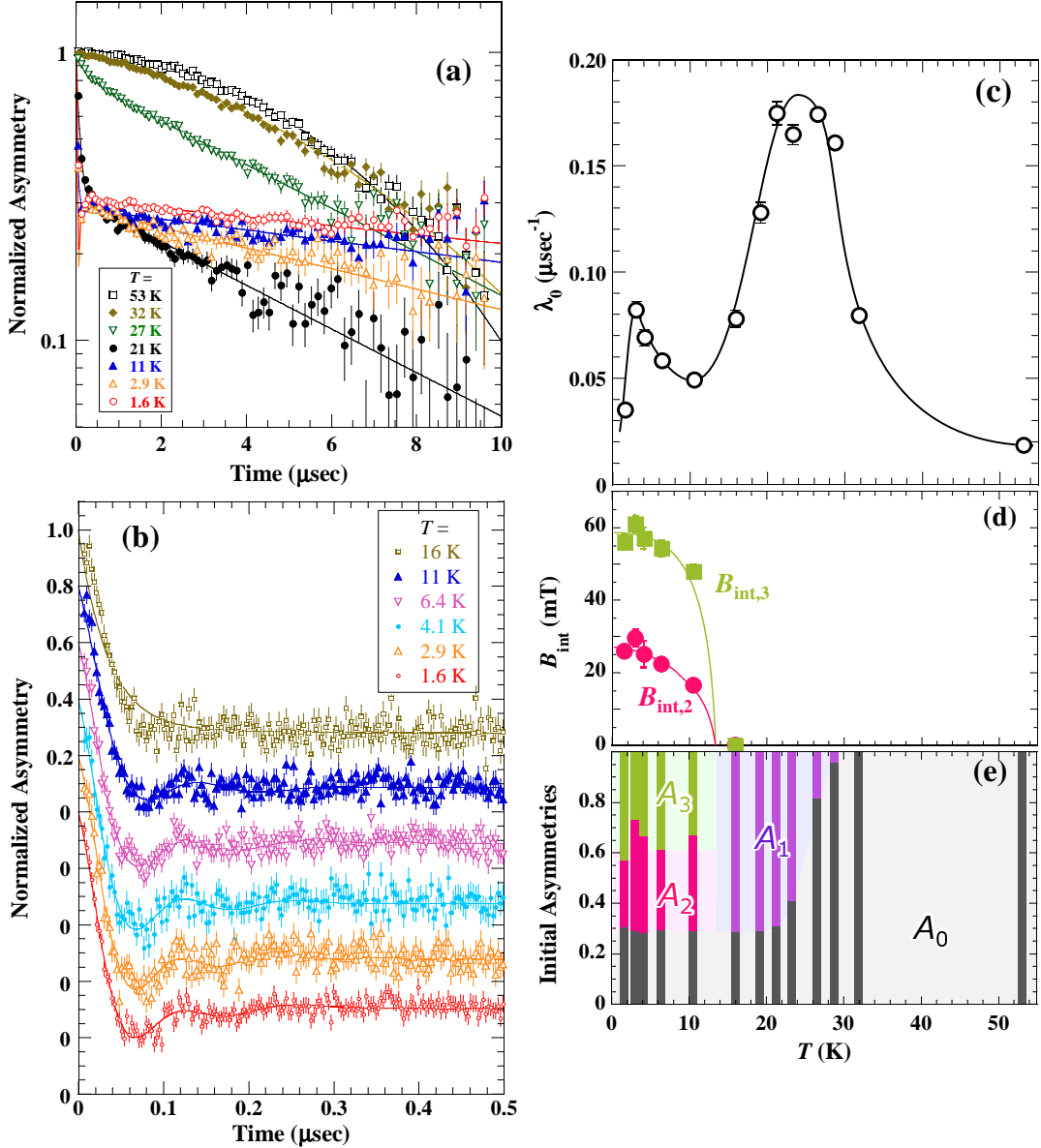}
			\caption{\label{musr2505}
				(Color online)
				(a)(b) Zero-field $\mu$SR time spectra of La$_{2-x}$Sr$_x$Cu$_{1-y}$Fe$_y$O$_4$ (LSCFO) with $x = 0.25$ and $y = 0.05$ ($p = 0.20$).
				The spectra in (b) are in the early time region at low temperatures and shifted upward regularly by 0.2 for clarity.
				All the spectra are shown after the subtraction of the background from the raw spectra and the normalization by the value at $t = 0$.
				Solid lines are the best-fit results using Eq. (\ref{eq1}).
				(c)(d)(e) Temperature dependences of (c) the depolarization rate of the first component in Eq. (\ref{eq1}), $\lambda _0$, (d) the internal magnetic fields, $B_\mathrm{int,2}$ and $B_\mathrm{int,3}$, and (e) the initial asymmetries of the respective components in Eq. (\ref{eq1}) for LSCFO with $x = 0.25$ and $y = 0.05$ ($p = 0.20$), respectively.
				Solid lines in (c)(d) and the colored areas in (e) are guides to the eye.
			}
		\end{figure*}
			The successive magnetic transitions have been confirmed also from the $\mu$SR measurements.
			Figure \ref{musr2505}(a) shows the ZF-$\mu$SR time spectra of LSCFO with $x = 0.25$ and $y = 0.05$ ($p = 0.20$) after the subtraction of the background from the raw spectra and the normalization by the value at $t = 0$.
			At a high temperature of 53 K, it is found that the spectrum shows a slow muon-spin-depolarization of the Gaussian-type due to nuclear dipole fields randomly oriented at the muon site,
			indicating that electron spins fluctuate fast beyond the $\mu$SR time window of $10^6$ -- $10^{11}$ Hz.
			With decreasing temperature, the muon-spin depolarization becomes fast progressively down to 21 K due to the development of the spin correlation.
			Focusing on the time region over $\sim$ 1 $\mu$sec, an upward shift of the spectrum with decreasing temperature is observed at low temperatures below 21 K.
			This is a behavior characteristic of the progress toward the formation of a static magnetic state.
			Remarkable is that the muon-spin depolarization is developed again at low temperatures below 11 K, resulting in the lower asymmetry at 2.9 K than that at 11 K.
			This means that spins fluctuate at 2.9 K faster than at 11 K.
			Finally, the spectrum shifts upward at 1.6 K, indicating the progress toward the formation of another static magnetic state.
			That is, the $\mu$SR spectra suggest the occurrence of successive magnetic transitions.
			
			Figure \ref{musr2505}(b) shows the data in the early time region at low temperatures, which are shifted upward regularly for clarity.
			An oscillatory behavior due to the formation of a static magnetic order is observed at low temperatures below 11 K, corresponding to the upward shift of the spectrum with decreasing temperature at low temperatures below 21 K in Fig. \ref{musr2505}(a).
			Carefully watching, the spectra below 11 K appear to consist of two components of oscillation: an apparant oscillation with the period of $\sim 0.13$ $\mu$sec and a slower oscillation damped fast with the period twice as large as the first one.

			To see more clearly, the $\mu$SR spectra were analyzed using the following equation:
			\begin{eqnarray}
					A(t) &=& A_0 e^{-\lambda _0t}G_\mathrm{KT}(\Delta ,t) +
					A_1e^{-\lambda _1t}\nonumber \\
					&&+ \sum_{i=2,3}[A_ie^{-\lambda _it}\cos (\gamma _\mu B_{\mathrm{int},i}t + \phi_i)].
				\label{eq1}
			\end{eqnarray}
			The first term represents the non-magnetic component in a region where electron spins fluctuate faster than the $\mu$SR time window at high temperatures.
			Note that the first term represents the so-called 1/3 component at low temperatures below the magnetic transition temperature.\footnote{The $\Delta$ in Eq. (\ref{eq1}) becomes zero, i.e. $G_\mathrm{z}(\Delta ,t) = 1$, at low temperatures below 30 K, because the effect of the nulcear-dipole field is masked by the effect of electron spins whose correlation is developed at low temperatures.}
			The second term represents the magnetically fluctuating component in a region where the electron-spin fluctuations slow down leading to a fast muon-spin-depolarization.
			The third and fourth terms represent two different magnetically ordered components in regions where static magnetic orders are formed.
			The $A_0$, $A_1$, $A_2$ and $A_3$ are initial asymmetries of the respective components, corresponding to respective volume fractions in a sample.
			The $G_\mathrm{KT}(\Delta ,t)$ is the static Kubo-Toyabe function describing the muon-spin depolarization due to nuclear dipole fields randomly oriented at the muon site with the distribution width, $\Delta$.
			The $\lambda _0$, $\lambda _1$, $\lambda _2$ and $\lambda _3$ are depolarization rates of the respective components.
			The $B_\mathrm{int,2}$ and $B_\mathrm{int,3}$ are internal magnetic fields at two different muon-sites and $\phi _2$ and $\phi _3$ are phases of the muon-spin precession.
			The $\gamma _\mu$ ($= 2\pi \times 135.5$ MHz/T) is the gyromagnetic ratio of muon spins.
			The $\mu$SR spectra are well fitted using Eq. (\ref{eq1}), as shown by solid lines in Fig. \ref{musr2505}(a) and (b).
			
			Figures \ref{musr2505}(c), (d) and (e) show the temperature dependences of $\lambda _0$, $B_\mathrm{int,2}$, $B_\mathrm{int,3}$, $A_0$, $A_1$, $A_2$ and $A_3$ of $x = 0.25$ and $y = 0.05$ ($p = 0.20$).
			Since it is well known that $\lambda _0$ exhibits a peak at the magnetic transition temperature, two peaks in Fig. \ref{musr2505}(c) indicate the occurrence of successive magnetic transitions.
			The temperature at which $\lambda _0$ shows a local maximum is defined as $T_{\lambda}$ and plotted in Fig. \ref{chiT0.20}(b).
			Obviously, the higher and lower values of $T_\lambda$ correspond to $T_\mathrm{g1}$ and $T_\mathrm{g2}$, respectively.
			Thus, the successive magnetic transitions observed in both $\mu$SR and magnetic susceptibility measurements are regarded as an intrinsic nature of overdoped LSCFO.

			As shown in Fig. \ref{musr2505} (d), two components of oscillation are developed almost simultaneously, so that $B_\mathrm{int,2}$ and $B_\mathrm{int,3}$ are $\sim 25$ mT and $\sim 60$ mT at 1.6 K.
			The former is comparable with the internal field observed in the underdoped regime of 1\% Fe-substituted LSCFO,\cite{KMS2012} while the latter is much larger than this.
			As shown in Fig. \ref{musr2505}(e), volume fractions of the two components, namely, $A_2$ and $A_3$ are comparable with each other.
			Moreover, it is found that neither the internal fields nor the initial asymmetries exhibit significant change at $T_\mathrm{g2}$.
			Accordingly, it is hard to attribute the two components of oscillation neither to the successive magnetic phases nor to an impurity phase.
			Since recent remeasurements in the antiferromagnetically ordered state of La$_2$CuO$_4$ with high statistics have uncovered minor muon-stopping sites in addition to the well-known major one,\cite{Watanabe_unpublish} the presence of two comparable muon-stopping sites may have to be taken into account in LSCFO.

			It is unexpected that an oscillation (two components of oscillation in detail) has been observed in the $\mu$SR spectra in the spin-glass state of 5\% Fe-substituted LSCFO at low temperatures below $T_\mathrm{g1}$, as shown in Fig. \ref{musr2505} (b), because no oscillation has been observed in the spin-glass state of 1\% Fe-substituted LSCFO in the overdoped regime due to the large distribution of the internal magnetic field as described in Sec. \ref{sec:introduction}.\cite{KMS2012proc,KMS2012}
			The oscillation may be due to the large amount of Fe$^{3+}$ spins in 5\% Fe-substituted LSCFO, which makes the internal-field distribution narrow.
			The SDW order of itinerant spins suggested from the elastic neutron-scattering experiment\cite{He2011PRL} may contribute the oscillation.

	\subsection{Hole-concentration dependence of the magnetic susceptibility in 5\% Fe-substituted La$_{2-x}$Sr$_x$Cu$_{1-y}$Fe$_y$O$_4$}
\begin{figure}[t]
				\includegraphics[width=0.70\linewidth]{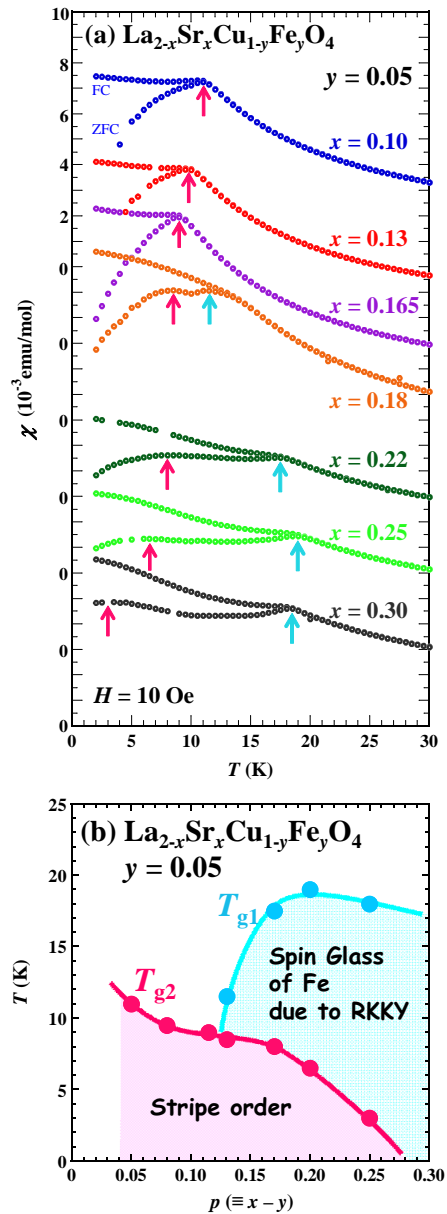}
				\caption{\label{chiTFe05}
					(Color online)
					(a) Temperature dependences of the field-cooled (FC) and zero-field-cooled (ZFC) magnetic susceptibility in a magnetic field of 10 Oe for La$_{2-x}$Sr$_x$Cu$_{1-y}$Fe$_y$O$_4$ (LSCFO) with $x = 0.10$ -- $0.30$ and $y = 0.05$.
					The data are shifted upward regularly  by $3 \times 10^{-3}$ emu/mol for clarity.
					Arrows represent $T_\mathrm{g1}$ or $T_\mathrm{g2}$ defined as the temperature at which the ZFC data show a local maximum.
					(b) Hole-concentration per Cu, $p$, dependences of $T_\mathrm{g1}$ and $T_\mathrm{g2}$ for LSCFO with $y = 0.05$.
					Solid lines are guides to the eye.
				}
			\end{figure}

			In order to investigate the evolution of the successive magnetic transitions with changing $p$, the magnetic susceptibility has been measured for 5\% Fe-substituted LSCFO in a wide range of $p$ from the underdoped regime to the overdoped one.
			Figure \ref{chiTFe05}(a) shows the temperature dependence of the magnetic susceptibility for LSCFO with $x = 0.10$ -- $0.30$ and $y = 0.05$.
			It is found that the successive magnetic transitions appear only in the overdoped regime of $x \ge 0.18$ ($p \ge 0.13$), while the hysteresis behavior is observed for all the samples.
			Hole-concentration dependences of $T_\mathrm{g1}$ and $T_\mathrm{g2}$ for 5\% Fe-substituted LSCFO are shown in Fig. \ref{chiTFe05}(b).
			Considering the evolution of $T_\mathrm{g1}$ and $T_\mathrm{g2}$ with changing $p$, the observed magnetic transition in the underdoped regime appears to correspond to $T_\mathrm{g2}$.
			Accordingly, it seems that $T_\mathrm{g2}$ decreases gradually with increasing $p$, while $T_\mathrm{g1}$ suddenly appears at $p \sim 0.13$ and keeps high in the overdoped regime.
			These results are consistent with the interpretation that a canonical spin-glass state of Fe$^{3+}$ spins due to the RKKY interaction is realized in the magnetic phase at low temperatures below $T_\mathrm{g1}$.\cite{KMS2012}
			This is because the RKKY interaction holds good in a metallic state and, in fact, conduction electrons in the overdoped regime are more itinerant than in the underdoped regime.
			As for the magnetic phase at low temperatures below $T_\mathrm{g2}$, the evolution of $T_\mathrm{g2}$ with changing $p$ suggests that the magnetic order below $T_\mathrm{g2}$ in the overdoped regime is the stripe order, because the magnetic order below $T_\mathrm{g2}$ in the underdoped regime is well known to be the stripe order.
			This means that the character of Cu spins changes from itinerant to localized at $T_\mathrm{g2}$ with decreasing temperature in the overdoped regime.
			Here, it is noted that the typical spin-glass behavior observed in the underdoped regime of $x \le 0.165$ ($p \le 0.115$) is inferred to be due to the formation of stripe-ordered clusters which are randomly oriented in zero field, as in the case of lightly-doped LSCO.\cite{Chou1995,Wakimoto2000}

	\subsection{Hole-concentration dependence of $\mu$SR in 1\% Fe-substituted La$_{2-x}$Sr$_x$Cu$_{1-y}$Fe$_y$O$_4$ in the overdoped regime}
		
		\begin{figure*}[t]
			\includegraphics[width=0.85\linewidth]{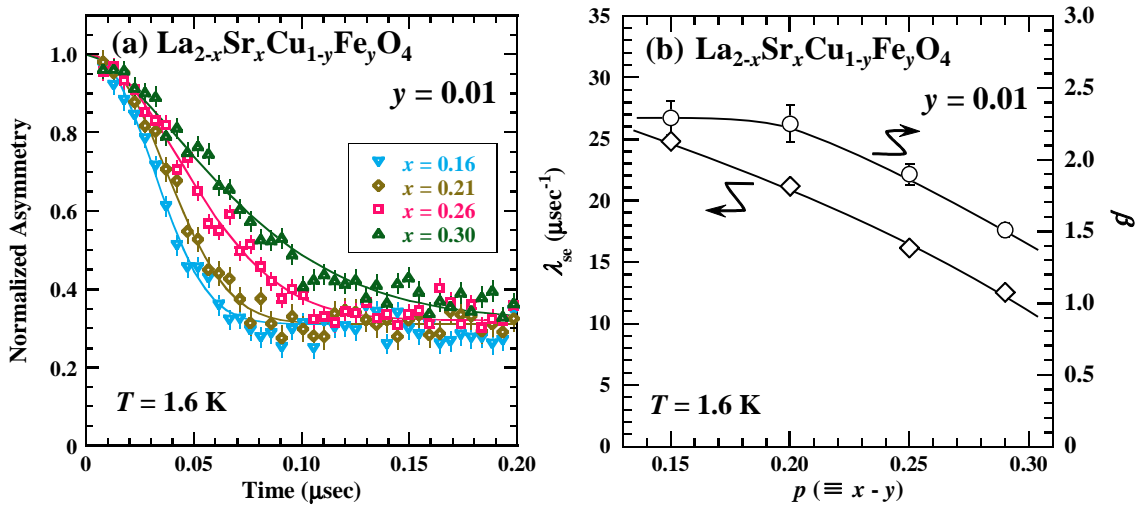}
			\caption{\label{musrFe01}
				(Color online)
				(a) Zero-field $\mu$SR time spectra at the measured lowest temperature of 1.6 K for La$_{2-x}$Sr$_x$Cu$_{1-y}$Fe$_y$O$_4$ with $x = 0.16$ -- $0.30$ and $y = 0.01$.
				All the spectra are shown after the subtraction of the background from the raw spectra and the normalization by the value at $t = 0$.
				Solid lines are the best-fit results using Eq. (\ref{eq2}).
				(b) Hole-concentration per Cu, $p$, dependences of $\lambda _\mathrm{se}$ and $\beta$ obtained from the best fit in (a) using Eq. (\ref{eq2}).
			}
		\end{figure*}
		
			In order to investigate the nature of the spin-glass state below $T_\mathrm{g1}$ in the overdoped regime in detail, $\mu$SR measurements have been carried out for 1\% Fe-substituted LSCFO with $x = 0.16$ -- $0.30$ and $y = 0.01$.
			Figure \ref{musrFe01}(a) shows the ZF-$\mu$SR time spectra at 1.6 K after the subtraction of the background from the raw spectra and the normalization by the value at $t = 0$.
			It is found that the so-called 1/3 tail of the $\mu$SR spectrum is observed for all the samples, indicating the formation of a static magnetic state.
			No oscillation is observed even in the $\mu$SR measurements using the dc muon beam with a high time-resolution, indicating that the internal magnetic field at the muon site is significantly distributed in magnitude.
			This is consistent with the results that the magnetic state below $T_\mathrm{g1}$ is a canonical spin-glass state.
			In fact, similar $\mu$SR spectra have been observed in dilute magnetic alloys such as Au-Fe and Cu-Mn exhibiting typical canonical spin-glass behaviors.\cite{Uemura1985}
			
			It is, however, noticed that the spectra shown in Fig. \ref{musrFe01}(a) are a little distinct from those of dilute magnetic alloys.
			That is, the initial relaxation of the spectrum appears to be of the Gaussian-type rather than of the exponential-type which is generally observed in spin-glass states in dilute magnetic alloys.
			To see clearly, the present $\mu$SR spectra were analyzed using the following equation with the stretched exponential function:
			\begin{equation}
				A(t) = A_\parallel \exp (-\lambda _0t) + A_\perp \exp \{-(\lambda _\mathrm{se}t)^{\beta}\}.
			\label{eq2}
			\end{equation}
			The first term represents the so-called 1/3 component of muon spins which is parallel to the internal magnetic field at $t = 0$ and depolarizes on account of fluctuation of the internal magnetic field.
			The second term represents the remaining 2/3 component of muon spins which is perpendicular to the internal magnetic field at $t = 0$ and depolarizes fast on account of the large distribution of the internal magnetic field.
			The $A_\parallel$ and $A_\perp$ are initial asymmetries, and $\lambda _0$ and $\lambda _\mathrm{se}$ are depolarization rates of the respective components.
			The $\beta = 1$ ($\beta = 2$) corresponds to the case that the spin density is dilute (dense) in a sample.\cite{Yaouanc2011}
			The $\mu$SR spectra are well fitted using Eq. (\ref{eq2}), as shown by solid lines in Fig. \ref{musrFe01}(a).
			
			Figure \ref{musrFe01}(b) shows values of $\lambda _\mathrm{se}$ and $\beta$ obtained from the best fit.
			For all the samples of 1\% Fe-substituted LSCFO in the overdoped regime, $\beta = 1.5$ -- $2.3$, suggesting that the fast muon-spin-depolarization is not caused exclusively by dilute (1\%) Fe$^{3+}$ spins.
			Moreover, the muon-spin depolarization is apparently faster in the present samples of LSCFO than in Au-Fe with the same amount of Fe spins.\cite{Uemura1985}
			This also suggests an additional contribution to the muon-spin depolarization besides the dilute Fe spins.
			Plausible candidates are polarized itinerant spins in the SDW state revealed by the neutron scattering experiment.\cite{He2011PRL}
			The magnitude of the magnetic moment is estimated to be 0.13 $\mu _\mathrm{B}$/Cu for LSCFO with $x = 0.26$ and $y = 0.01$.
			This value is not tiny compared with 0.65 $\mu _\mathrm{B}$/Cu of La$_2$CuO$_4$\cite{Forsyth1988} and even larger than 0.1 $\mu _\mathrm{B}$/Cu of LSCO with $x = 0.12$.\cite{Kimura1999,Wakimoto2001}
			It is likely that the polarization of itinerant spins is not only due to the RKKY interaction but also be affected by the Fermi-surface nesting.\cite{He2011PRL}
			That is, it seems that the good nesting in overdoped LSCFO enhances the polarization of itinerant spins, so that a SDW order is stabilized to be observed in the neutron-scattering experiment\cite{He2011PRL} and contributes to the generation of the internal magnetic field at the muon site together with dilute localized Fe$^{3+}$ spins as described in Sec. \ref{sec:RAD1}.
			Therefore, the present results demonstrate the presence of a novel coexistent state of a SDW order and a spin-glass state in a metal including dilute localized spins and possessing a Fermi surface with a good nesting vector.
			
			It is found that the value of $\beta$ decreases with hole doping.
			This is interpreted as being due to the reduction of the contribution of itinerant spins to the muon-spin depolarization, which is responsible for the Gaussian-type relaxation, with hole doping and due to the alternatively dominating contribution of dilute Fe$^{3+}$ spins leading to the exponential-type relaxation.
			It is found that the value of $\lambda _\mathrm{se}$, which reflects the magnitude of the internal magnetic field at the muon site, decreases with hole doping, too.
			The reason why the contribution of itinerant spins, namely, the SDW order to the muon-spin depolarization is reduced with hole doping is not so clear at present.
			Taking into account the results that both $T_\mathrm{g1}$ in 5\% Fe-substituted LSCFO and the magnetic transition temperature, $T_\mathrm{N}$, of 1\% Fe-substituted LSCFO in our previous paper\cite{KMS2012} decrease with increasing $p$ in the overdoped regime, the weakening of the polarization of the itinerant spins, namely, the weakening of the electron correlation with increasing $p$ in the overdoped regime may be related to these results.

	\subsection{Phase diagram of La$_{2-x}$Sr$_x$Cu$_{1-y}$Fe$_y$O$_4$}
		
\begin{figure}[t]
				\includegraphics[width=0.95\linewidth]{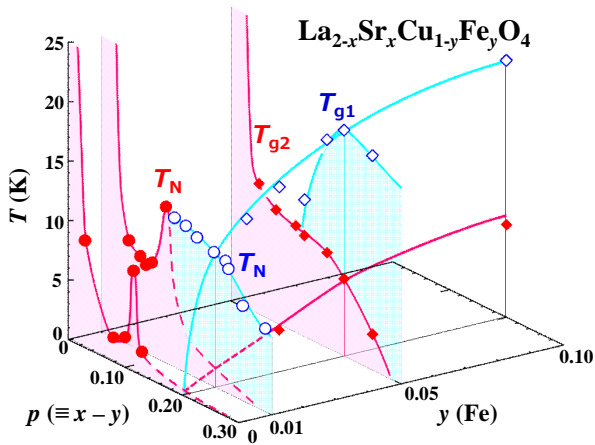}
				\caption{\label{3dpd}
					(Color online)
					Magnetic phase diagram in $p\ (\equiv x -y)$-$y$-$T$ space for La$_{2-x}$Sr$_x$Cu$_{1-y}$Fe$_y$O$_4$.
					Circles represent the magnetic transition temperature, $T_\mathrm{N}$, obtained in our former $\mu$SR measurements of LSCO\cite{Adachi2004Zn,Adachi2008,Tanabe2011} and LSCFO.\cite{KMS2012}
					Diamonds represent $T_\mathrm{g1}$ or $T_\mathrm{g2}$ at which the magnetic susceptibility in a magnetic field of 10 Oe shows a local maximum.
					Closed symbols represent long-range antiferromagnetic transition temperatures or stripe-ordered temperatures based on localized Cu$^{2+}$ spins.
					Open symbols represent spin-glass transition temperatures of Fe$^{3+}$ spins due to the RKKY interaction based on itinerant spins.
					Solid lines are guides for the eye.
					Broken lines are guides for the eye based on our speculations.
				}
			\end{figure}

			Figure \ref{3dpd} shows the obtained phase diagram of LSCFO together with the data of LSCO\cite{Adachi2004Zn,Adachi2008,Tanabe2011} and 1\% Fe-substituted LSCFO.\cite{KMS2012}
			The $T_\mathrm{N}$ for LSCO and 1\% Fe-substituted LSCFO was obtained in our former $\mu$SR measurements.\cite{KMS2012}
			For 5\% Fe-substituted LSCFO, the static magnetic state at low temperatures below $T_\mathrm{g2}$ in the overdoped regime is related to localized Cu$^{2+}$ spins observed in the underdoped regime, suggesting that the stripe correlations continue up to $p \sim 0.28$.
			As described in Sec. \ref{sec:RAD1}, it appears that the static magnetic state exists at low temperatures in Fe-free LSCO with $p = 0.20$, though it may be suppressed by the superconductivity.
			Therefore, it is inferred that the stripe correlations exist up to $p \sim 0.28$ even in LSCO.
			This inference is consistent with the $\mu$SR results of LSCZO that the Zn-induced development of the Cu-spin correlation is observed up to $p \sim 0.30$\cite{Risdiana2008} and with the inelastic neutron-scattering results that the incommensurate spin-correlation is observed up to $p \sim 0.30$.\cite{Wakimoto2007}
			Accordingly, it seems that the stripe correlations are closely related to the appearance of the high-$T_\mathrm{c}$ superconductivity.\cite{Kivelson2003}
			
\section{Summary}
			In order to investigate the relationship between the stripe correlations and the magnetism in the overdoped regime induced by the Fe substitution, we have carried out magnetic susceptibility and ZF-$\mu$SR measurements in LSCFO with various $x$ and $y$.
			Successive magnetic transitions have been observed in overdoped LSCFO.
			The magnetic phase at higher temperatures has been found to be a spin-glass state of Fe$^{3+}$ spins due to the RKKY interaction based on itinerant spins.
			The magnetic phase at lower temperatures has been found to be developed continuously to the underdoped regime with decreasing hole-concentration, suggesting that the static magnetic state is a stripe-ordered state of localized Cu$^{2+}$ spins.
			That is, it is likely that the character of Cu spins changes from itinerant to localized with decreasing temperature in the overdoped regime.
			Moreover, the static magnetic state at lower temperatures has been found to exist up to $p \sim 0.28$ where the superconductivity in pristine LSCO is terminated.
			Therefore, it seems that the stripe correlations are closely related to the high-$T_\mathrm{c}$ superconductivity.
			It is a future work to investigate the static magnetic state at lower temperatures in detail.

			As for the spin-glass phase, it has been found that muon spins depolarize not only by dilute Fe$^{3+}$ spins but also by polarized itinerant spins, contrasting to the usual muon-spin-depolarization in dilute alloys such as Au-Fe.\cite{Uemura1985}
			Taking into account the neutron-scattering results\cite{He2011PRL} that clear incommensurate magnetic peaks due to a SDW order have been observed in 1\% Fe-substituted LSCFO in the overdoped regime, the present results demonstrate the presence of a novel coexistent state of the SDW order and the spin-glass state owing to both dilute localized Fe$^{3+}$ spins and the good Fermi-surface-nesting.
			
\section*{Acknowledgments}
			We would like to thank R. Kadono and M. Sasaki for their helpful discussion.
			The $\mu$SR measurements at PSI were partially supported by the KEK-MSL Inter-University Program for Oversea Muon Facilities and also by the Global COE Program “Materials Integration (International Center of Education and Research), Tohoku University,” of the Ministry of Education, Culture, Sports, Science and Technology, Japan.
			One of the author (KMS) was supported by the Japan Society for the Promotion of Science.

			\bibliography{_FeLSCO_2nd}

\end{document}